\documentclass[sigconf, nonacm]{acmart}

\usepackage[noend]{algorithmic}
\usepackage{array,multirow}
\usepackage{xspace}
\usepackage{pifont}
\usepackage{adjustbox}
\usepackage{xcolor}
\usepackage{colortbl}
\usepackage{pbox}

\usepackage{amssymb}
\usepackage{graphicx}
\usepackage{arydshln}
\usepackage{subfigure}
\usepackage{bm}
\usepackage{amsmath}
\usepackage{makecell}
\usepackage{url}

\newcommand{\CE}{\textsf{CardEst}\xspace}
\newcommand{\Glue}{\textsc{Glue}\xspace}

\newcommand{\Card}{{\sf Card}\xspace}
\newcommand{\Dom}{{\sf Dom}\xspace}
\newcommand{\Smp}{{\sf Smp}\xspace}
\newcommand{\Dis}{{\sf Dis}\xspace}

\mathchardef\mhyphen="2D

\definecolor{mygrey}{RGB}{230,230,240}

\newcommand{\CEend}{\texttt{CardEst}}

\begin{document}
	\title{Glue: Adaptively Merging Single Table Cardinality \break
	to  Estimate Join Query Size}
\author{Rong Zhu$^1$, Tianjing Zeng$^{1, 2}$, Andreas Pfadler$^1$, Wei Chen$^1$, Bolin Ding$^1$, Jingren Zhou$^1$}
	\vspace{0.5em}
	\affiliation{%
	\institution{\LARGE{\textit{$^1$Alibaba Group}, \textit{$^2$Renmin University of China}} \\
	\textsf{\{red.zr, zengtianjing.ztj, andreaswernerrober, wickeychen.cw, bolin.ding,  jingren.zhou\}@alibaba-inc.com}
	}
	\vspace{0.5em}
	}

\begin{abstract}
Cardinality estimation (\CE), a central component of the query optimizer, plays a significant role in generating high-quality query plans in DBMS. The \CE problem has been extensively studied in the last several decades, using 
both traditional and ML-enhanced methods. Whereas, the hardest problem in \CE, i.e., how to estimate the join query size on multiple tables, has not been extensively solved. Current methods either reply on independence assumptions or apply techniques with heavy burden, whose performance is still far from satisfactory. Even worse, existing \CE methods are often designed to optimize one goal, i.e., inference speed or estimation accuracy, which can not adapt to different occasions.

In this paper, we propose a very general framework, called \Glue, to tackle with these challenges. Its key idea is to elegantly decouple the correlations across different tables and losslessly merge single table \CE results to estimate the join query size. \Glue supports obtaining the single table-wise \CE results using any existing \CE method and can process any complex join schema. Therefore, it easily adapts to different scenarios having different performance requirements, i.e., OLTP with fast estimation time or OLAP with high estimation accuracy.
Meanwhile, we show that \Glue can be seamlessly integrated into the plan search process and is able to support counting distinct number of values. All these properties exhibit the potential advances of deploying \Glue in real-world DBMS.
\end{abstract}

\maketitle

\section{Introduction}
Query optimizer (QO) plays a significant important role in modern DBMSs. It is an integral component to generate high-quality execution plans for the input SQL queries. \emph{Cardinality estimation(\CE)}, which aims at estimating the result size of all sub-plans queries, is a central part in QO. It lays the foundation for cost estimation and guides the QO for join order selection. Thus, \CE has a critical impact on the quality of the generated query plans.

\smallskip

\noindent{\underline{\textbf{Background:}}}
Due to its importance, \CE has been extensively studied in the literature. The core task of \CE is to build a compact sketch capturing the synopses of data and/or query information. Current open-source and commercial DBMSs mainly use two traditional \CE methods, namely histogram~\cite{selinger1979access,gunopulos2005selectivity,bruno2001stholes,muralikrishna1988equi,wang2003multi,deshpande2001independence} in PostgreSQL\cite{psql2020} and SQL Server\cite{sqlserver2019} and sampling~\cite{leis2017cardinality,heimel2015self,kiefer2017estimating,zhao2018random, li2016wander} in MySQL~\cite{mysql2020} and MariaDB~\cite{mdb2020}.
Recently, with the prosperity of machine learning (ML), there is a booming of ML-enhanced \CE methods in the last several years~\cite{kipf2018learned,hilprecht2019deepdb, sun2019end,yang2019deep,yang2020neurocard,wu2020bayescard,zhu2020flat,hasan2020,wu2021uae,dutt2019selectivity, liu2021fauce}. These methods are either query-driven~\cite{kipf2018learned, dutt2019selectivity}, which maps featurized queries to their cardinality, or data-driven, which directly model the joint distribution of all attributes~\cite{yang2019deep,yang2020neurocard, tzoumas2011lightweight, getoor2001selectivity,wu2020bayescard, hilprecht2019deepdb, zhu2020flat, liu2021fauce}.
They devote lots of efforts in improving the performance of \CE algorithms in terms of different criteria, namely end-to-end query time~\cite{han2021CEbenchmark}, estimation accuracy, inference latency, updating speed and model size~\cite{zhu2020flat,liu2021fauce}.

\smallskip

\noindent{\underline{\textbf{Challenge and Motivation:}}}
Although the estimation accuracy, as well as other performance criteria, has been shown to be significantly improved for \CE methods. Some challenges still exist for \CE algorithms. We summarize them as follows.

First, join query size estimation, the hardcore problem of \CE, is not well studied yet. Due to the booming size of the join table and the existence of cross-table correlations between attributes, the performance of existing \CE methods degrades with the number of join tables~\cite{han2021CEbenchmark}. Current \CE methods mainly apply two kinds of approaches to process join queries. The first one~\cite{yang2020neurocard,wu2021uae} directly builds a large model on (the samples of) the full outer join table, which causes heavy overhead and poor scalability. The second one~\cite{hilprecht2019deepdb,wu2020bayescard,zhu2020flat} makes strong independence assumptions among tables and builds an ensemble of small models on partial tables, which needs to be tuned by experience and would more or less harm the estimation accuracy.

Second, current \CE methods are not adaptive to different applications. 
Real-world DBMS would face different datasets and query workloads, which emphasize different performance criteria of \CE methods. For example, OLTP queries require fast estimation time while OLAP queries need high estimation accuracy~\cite{han2021CEbenchmark}. These criteria often conflicts with each other so it is difficult for a \CE method to achieve both at the same time. Moreover, even for a specific metric, different \CE methods are suitable for different data due to their independence assumptions. For example, SPN-based method~\cite{hilprecht2019deepdb} is only accurate on attributes with low correlations. There exist no versatile \CE method that can perform well on any data with any query workload.

As a result, a more sophisticated and universal \CE paradigm is still missing, especially for  the complex multi-table join queries.

\smallskip
\noindent{\underline{\textbf{Contributions:}}}
In this paper, we propose \Glue, a novel \CE framework to tackle with these challenges.
Unlike with existing \CE methods, \Glue does not consider how to model data in each single table. Instead, it builds an upper structure that decouples the correlations between join tables using local independence, and then merge single table \CE results to predict join query size. The upper structure is conceptually independent of the underlying models. In comparison with existing \CE methods, \Glue has the following advantages:

1) it is very general to support any join schema, i.e. star, chain, cycle and mixture, and join types, i.e., one-to-many and many-to-many. The upper structure is very lightweight and easy to update. Meanwhile, \Glue's local independence assumption is data adaptive, so its estimation error is much lower.

2) it is very flexible to support any \CE method on single table as plug-ins, or even different \CE methods for different tables in one database. This allows the QO to steer to different optimization goals, i.e. plan quality or throughput, and adapt to different types of data, i.e. loosely or strongly correlated, by selecting proper \CE methods. To the best of our knowledge, this establish a new paradigm for \CE.

Besides, we show that the computation process in \Glue ensembles the dynamic programming based join order selection in QO, so it can be seamlessly integrated to speed up the plan search process. More over, we show that \Glue could also support counting distinct values. 
All these properties indicate that \Glue is a highly promising candidate for deploying in real-world DBMS.

\smallskip
\noindent{\underline{\textbf{Organization:}}}
In the following content, Section~2 introduces some preliminary knowledge, Section~3 describes the main idea of \Glue framework, Section~3 presents how to construct the structure in \Glue, Section~4 exhibits the distinct counting method in \Glue and Section~5 concludes this paper.


\section{Preliminaries}
\label{sec: prelim}

In this section, we formalize the \CE problem and brief review representative \CE algorithms.

\subsection{Problem Definition}

Let $T$ be a table with $k$ attributes $A = \{A_1, A_2, \dots, A_k \}$. $T$ could either be a single relational table or a joined table. Without ambiguity, if $T$ is a joined table, we also use $T$ to represent the set of all single tables joining it. In this paper, we assume that each attribute $A_i$ for each $1 \leq i \leq k$ to be either categorical (whose values can be mapped to integers) or continuous, whose domain (all unique values) is denoted as $\Dom(A_i)$. We also denote $\Dom(T) = \Dom(A_1) \times \Dom(A_2) \times \dots \times \Dom(A_k)$ to be the domain of table $T$.

Thereafter, any selection query $Q$ on $T$ can be represented in a canonical form: $Q = \{A_1 \in R_1 \wedge A_2 \in R_2 \wedge \cdots \wedge A_n \in R_n\}$, where $R_i \subseteq \Dom(A_i)$ is the constraint region specified by $Q$ over attribute $A_i$ (i.e. filter predicates). Without loss of generality, we have $R_i =  \Dom(A_i)$ if $Q$ has no constraint on $A_i$. In this paper, if a region $R \subseteq \Dom(T)$ could be decomposed into the form $R = R_1 \times R_2 \times \dots \times R_k$ where $R_i \subseteq \Dom(A_i)$ for all $i$, we call $R$ is a \emph{regular} region. Intuitively, a regular region is formed by a number of hyper-rectangles in the domain space. Obviously, the query space of any selection query $Q$ is a regular region. In the following, we use $Q$ to denote the selection query and its region interchangeably. In this paper, we focus on evaluating selection queries on numerical or categorical attributes. We do not consider `LIKE'' (or pattern matching) queries on string attributes as they follow different technical routines.

Let $\Card(T, Q)$ denote the \emph{cardinality}, i.e., the exact number of records in $T$ satisfying all constraints in $Q$. The \CE problem requires estimating $\Card(T, Q)$ as accurately as possible without executing $Q$ on $T$. 

The \CE problem is often interpreted and solved in a statistical perspective. Specifically, we could regard each attribute $A_i$ as a random variable defined over its domain space $\Dom(A_i)$. Then, the set of attributes $A$ defines a joint probability distribution function (PDF) $\Pr_{T}(A) = \Pr_{T}(A_1, A_2, \dots, A_k)$ over table $T$. Each record $t \in T$ represents an independent tuple sampled from $\Pr_{T}(A)$.

At this time, $\Pr_{T}(Q) = \Pr_{T}(A_1 \in R_1, A_2 \in R_2, \dots, A_k \in R_k)$ represents the probability that a randomly picked record $t \in T$ satisfying the query $Q$. When the number of tuples is large enough in $T$, we naturally have $\Card(T, Q) = \Pr_{T}(Q) \cdot |T|$. When $T$ is a join table, e.g. $T \bowtie S$, we have $\Card(T \bowtie S, Q) = \Pr_{T \bowtie S}(Q) \cdot |T \bowtie S|$.
Since $|T|$ or $|T \bowtie S|$ is often known or can be estimated easily, the \CE problem is equivalent to model the joint PDF $\Pr_{T}(A)$ and estimate the probability $\Pr_{T}(Q)$.

\subsection{Related Work}
There exist many \CE methods in the literature, which can be classified into three classes as follows:

\emph{Traditional \CE methods}, such as histogram~\cite{selinger1979access} and sampling~\cite{leis2017cardinality,heimel2015self,kiefer2017estimating}, are widely applied in DBMS and generally based on simplified assumptions and expert-designed heuristics. 
Many variants of histograms~\cite{poosala1997selectivity, deshpande2001independence, gunopulos2000approximating, gunopulos2005selectivity, muralikrishna1988equi, wang2003multi, bruno2001stholes, srivastava2006isomer, khachatryan2015improving, fuchs2007compressed, stillger2001leo, wu2018towards} are proposed later to enhance their performance. Sampling-based variants include query-driven kernel-based methods~\cite{heimel2015self,kiefer2017estimating}, index based  methods~\cite{leis2017cardinality} and random walk based  methods~\cite{zhao2018random, li2016wander}.
Some other work, such as the sketch based method~\cite{cai2019pessimistic}, explores a new direction for \CEend.

\textit{ML-based query-driven \CE methods} try to learn a model to map each featurized query $Q$ to its cardinality $\Card(T, Q)$ directly. Some ML-enhanced methods improve the performance of \CE methods by using more complex models such as DNNs~\cite{kipf2018learned} or gradient boosted trees~\cite{dutt2019selectivity}.

\textit{ML-based data-driven \CE methods} are independent of the queries. They try to model the joint PDF $\Pr_{T}(A)$ directly so they have better generalization ability. A variety of ML-based models have been used in existing work to represent $\Pr_T(A)$, the most representative of which includes deep auto-regression model~\cite{yang2019deep,yang2020neurocard,hasan2019multi} and probabilistic graphical models (PGMs) such as
Bayesian networks (BN)~\cite{tzoumas2011lightweight, getoor2001selectivity, wu2020bayescard}, SPN~\cite{hilprecht2019deepdb}, and FSPN~\cite{zhu2020flat}. 
In addition, some methods proposed recently such as~\cite{wu2021uae} try to integrate both query and data information for \CE.


\section{\Glue Framework}
\label{sec: method}

In this section, we describe the details of our \Glue framework. We first formally define local independence, the foundation tool for our \Glue framework in Section~\ref{sec: method-1}, then present the main idea of \Glue in Section~\ref{sec: method-2}. Finally, Section~\ref{sec: method-3} shows how \Glue is applied to solve the \CE problem.

\begin{figure*}[!t]
\centering
	\begin{equation}
	\label{eq: prtqxqy}
	\begin{split}
	\Pr\nolimits_{T}(Q) & = \sum_{q \in Q} \Pr\nolimits_{T}(q_{Y}) \cdot \Pr\nolimits_{T}(q_{X} | q_{Y}) = \sum_{i} \sum_{q \in Q \cap (\Dom(X) \times R_i) } \Pr\nolimits_{T}(q_{Y}) \cdot \Pr\nolimits_{T}(q_{X} | q_{Y})  = \sum_{i} \sum_{q \in Q \cap (\Dom(X) \times R_i) } \Pr\nolimits_{T[R_i]}(q_{Y}) \cdot \Pr\nolimits_{T[R_i]}(q_{X}) \\
	& = \sum_{i} \left( \left(\sum_{q \in Q \cap (\Dom(X) \times R_i) } \Pr\nolimits_{T[R_i]}(q_{Y}) \right) \cdot \left(\sum_{q \in Q \cap (\Dom(X) \times R_i) } \Pr\nolimits_{T[R_i]}(q_{X}) \right)\right)  = \sum_{i} \Pr\nolimits_{T[R_i]}(Q_{Y}) \cdot \Pr\nolimits_{T[R_i]}(Q_{X} \cap R_i))).
	\end{split}
\end{equation}
\end{figure*}

\subsection{Local Independence}
\label{sec: method-1}

We first introduce the concept of local independence, which serves as a fundamental tool in our \Glue framework. For any table $T$ having attributes (random variables) $A$, let $X$ and $Y$ be a division of $A$. For any query $Q$ on $T$, we have 
\begin{equation}
\label{eq: prtq}
\Pr\nolimits_{T}(Q) = \sum\nolimits_{q \in Q} \Pr\nolimits_{T}(q_{Y}) \cdot \Pr\nolimits_{T}(q_{X} | q_{Y}),
\end{equation}
where $q_{X}$ and $q_Y$ represent the values of the point $q$ on attributes $X$ and $Y$, respectively. The hardness in Eq.~\eqref{eq: prtq} is that the term $\Pr_{T}(q_{X} | q_{Y})$ is not independent of $q_{Y}$, i.e., the conditional PDF $\Pr_{T}(X|y)$ differs for different value $y$ of $Y$, so the probability over $X$ and $Y$ can not be computed and then multiplied independently. We derive a method to decouple the correlations between $X$ and $Y$.

To compactly model the conditional PDF $\Pr(X|Y)$, we partition the domain space $\Dom(T) = \Dom(X) \times \Dom(Y)$ into multiple regions in terms of $Y$ as $\Dom(X) \times R_1, \Dom(X) \times R_2, \dots, \Dom(X) \times R_k$. Each $R_i \subseteq \Dom(Y)$ is a regular subspace s.t.~for any $y, y' \in R_i$, $\Pr_{T}(X|y) = \Pr_{T}(X|y')$ roughly holds. At this time,
we only need to maintain the PDF $\Pr_{T[R_i]}(X)$ for each $R_i$, where $T[R_i]$ denote the set of tuples in $T$ existing in the space $R_i$. We have $\Pr_{T}(X|y) = \Pr_{T[R_i]}(X)$ for any $y \in R_i$. We call this the \emph{contextual condition removal}, where each sub-domain $R_i$ refers to the context.

For each sub-space $\Dom(X) \times R_i$, we have $\Pr_{T}(Y) =  \Pr_{T[R_i]}(Y)$ when the value of $Y$ is restricted in $R_i$. Let $Q_X$ and $Q_Y$ denote the space of $Q$ restricted to the domain space of $X$ and $Y$, respectively. Then, we derive Eq.~\eqref{eq: prtqxqy}, which splits the probability of $X$ and $Y$ to be independent terms in each region $Q \cap (\Dom(X) \times R_i)$. Using this local independence, we could fast compute the probability of query $Q$ in each region and then sum them together.

Next, we introduce the intuitive idea on how to divide the domain space to derive local independence. 
In the extreme case, the local independence for each distinct value in $Dom(T)$. In general, we could break the correlations between variables $X$ and $Y$ to derive the local independence. As computing the correlations of two set of variables are expensive, we could done in a pairwise manner using a heuristic rule. Specifically, we could compute the pairwise correlation value $s(X_i, Y_j)$, such as RDC score~\cite{lopez2013randomized}, for each pair of attributes $X_i \in X$ and $Y_j \in Y$ and select the $Y_j$ maximizing $s(X_i, Y_j)$, i.e. $Y_j = \arg\max_{j} s(X_i, Y_j)$. Then we could divide the domain according to $Y_j$ by splitting its domain into several parts. After partition, all tuples in the same part
tend to have more similar values in terms of $Y_j$, so $Y_j$ is not easily affected by $X$ and more likely to be locally independent of $X$. In an extreme case, all records in the same part may have the same value on $Y_j$ and obviously $Y_j$ is independent of $X$ in this part. This method could be iteratively done over each part until the maximum pairwise correlation value $s(X_i, Y_j)$ is lower than a threshold.

\subsection{Overview of \Glue}
\label{sec: method-2}

In this subsection, we generally introduce how \Glue do \CE on join tables in a top-down manner.
Let $W = T \bowtie S$ be a join relation where $T = (A_1, A_2, \dots, A_k)$ and $S = (B_1, B_2, \dots, B_n)$ represent a single or join relation table.
For any selection query $Q$ on $T \bowtie S$, let $Q_T$ and $Q_S$ denote the sub-query on  $T$ and $S$, respectively. By Section~\ref{sec: prelim}, we have $\Card(T \bowtie S, Q) = \Pr_{T \bowtie S}(Q) \cdot |T \bowtie S|$. \Glue could obtain $\Pr_{T \bowtie S}(Q)$ by merging the information from $T$ and $S$ without building the entire model over $T \bowtie S$. We elaborate the details as the following three main steps.

\smallskip

\noindent \underline{\textbf{Step~1: Table decoupling.}}
Note that, some attributes $A_i$ of $T$ may be correlated with some $B_j$ of $S$ over $W = T \bowtie S$, which is also known as the cross-table correlations. Therefore, we often have $\Pr_{W}(Q) \neq  \Pr_{T \bowtie S}(Q_T)  \Pr_{T \bowtie S}(Q_S)$ on the whole domain space of $Dom(T) \times Dom(S)$.
At this time, we utilize the local independence tool, namely \emph{cross-table local independence}, to  decompose $\Pr_{W}(Q)$.

Specifically, let $L_1, L_2, \dots, L_t$ be the partition of $\Dom(T) \times \Dom(S)$ and each $L_i$ is in a regular form $L_i = \Dom(T) \times L^{1}_i \times L^{2}_i \times \dots \times L^{i}_t$ with $L^{j}_i \subseteq \Dom(B_j)$ for each $j$. The local independence holds in each $L_i$. That is, we have
\begin{equation*}
 \Pr\nolimits_{T \bowtie S}(Q \cap L_i) = \Pr\nolimits_{W}(Q \cap L_i) = \Pr\nolimits_{W[L_i]}(Q_T) \cdot \Pr\nolimits_{W[L_i]}(Q_S \cap L_i)
\end{equation*}
for each $L_i$ by Eq.~\eqref{eq: prtqxqy}. Then, using  Eq.~\eqref{eq: prtqxqy}, we derive

\begin{equation}
\label{eq: TSsplit}
\begin{split}
\Pr\nolimits_{T \bowtie S}(Q) & = \Pr\nolimits_{W}(Q)  =  \sum_{i} \Pr\nolimits_{W}(Q \cap L_i) \\
& = \sum_{i} \Pr\nolimits_{W[L_i]}(Q_T) \cdot  \Pr\nolimits_{W[L_i]}(Q_S \cap L_i).
\end{split}
\end{equation}
At this time, on each $L_i$, we could separately compute the probability of $Q_T$ and $Q_S \cap L_i$.
As $Q_T$ and $Q_S \cap L_i$ contains no predicates over another table, we decouple the connections across the tables $T$ and $S$.  We next show how to obtain the probability purely using the models on $T$ and $S$, respectively. 

\smallskip
\noindent \underline{\textbf{Step~2: Local PDF correction.}}
The probability $\Pr_{W[L_i]}(Q_S \cap L_i)$ contains only predicates on attributes in $S$, but it is defined over the joint PDF of $W = T \bowtie S$ but not over $S$. We apply the idea proposed in~\cite{zhu2020flat, hilprecht2019deepdb} to correct the PDF from $S$ to $T \bowtie S$ using fanout(or scattering) columns. However, unlike with them, we do not need to explicitly maintain these  fanout columns but just need to store several numbers. Let $F_{S \to T}$ denote the fanout column from table $S$ to $T$. For any tuple $t$ in $S$, $F_{S \to T}(t)$ records how many tuples in $T$ could join with $t$. Notably, we set $F_{S \to T}(t) = 1$ if $t$ can not join with any tuple in $T$. At this time, $t$ still occurs once (with \textsf{null} attributes) in $T \bowtie S$. Then, by~\cite{zhu2020flat, hilprecht2019deepdb}, we have
\begin{equation}
\Pr\nolimits_{T \bowtie S}(Q_S \cap L_i) = \frac{|S|}{|T \bowtie S|} \cdot \sum_{f}
\Pr\nolimits_{S}(Q_S \cap L_i) \wedge F_{S \to T} = f) \cdot f.
\end{equation}
The above equation is difficult to compute directly. However, we could simplify it again using the local independence property. We call such the \emph{join-key frequency independence}.

This time we divide the domain space $\Dom(S)$ to regular sub-space $S_1, S_2, \dots S_d$ s.t. all attributes of $S$ are locally independent of $F_{S \to T}$ in each $S_j$. That is, for any space $Q' \in \Dom(S)$, we have 
\begin{equation*}
\Pr\nolimits_{S}((Q' \cap S_j) \wedge F_{S \to T} = f) = \Pr\nolimits_{S[S_j]}(Q' \cap S_j)  \cdot \Pr\nolimits_{S[S_j]}(F_{S \to T} = f).
\end{equation*}
Then, let $Q'_i = Q_S \cap L_i$. By  Eq.~\eqref{eq: prtqxqy}, we derive
\begin{equation}
\label{eq: Ssplit}
\begin{split}
& \Pr\nolimits_{T \bowtie S}(Q'_i) \\
& = \frac{|S|}{|T \bowtie S|} \cdot 
\sum_{j} \sum_{f} \Pr\nolimits_{S}((Q'_i \cap S_j) \wedge F_{S \to T} = f) \cdot f \\
& = \frac{|S|}{|T \bowtie S|} \cdot 
\sum_{j} \sum_{f} \Pr\nolimits_{S[S_j]}(Q'_i \cap S_j)  \cdot \Pr\nolimits_{S[S_j]}(F_{S \to T} = f) \cdot f \\
& = \frac{|S|}{|T \bowtie S|} \cdot 
\sum_{j} \left( \Pr\nolimits_{S[S_j]}(Q'_i \cap S_j) \cdot \sum_{f} \Pr\nolimits_{S[S_j]}(F_{S \to T} = f) \cdot f \right) \\
& = \frac{|S|}{|T \bowtie S|} \cdot 
\sum_{j} \Pr\nolimits_{S[S_j]}(Q'_i \cap S_j) \cdot \mathbb{E}_{S[S_j]}[F_{S \to T}],
\end{split}
\end{equation}
where $\mathbb{E}_{S[S_j]}[F_{S \to T}]$ denotes the expected value of $F_{S \to T}$ restricted to all tuples of $S$ in the sub-domain $S_j$.
We could pre-obtain and store all these expected values. The details are discussed later. All 
probabilities $\Pr\nolimits_{S[S_j]}(Q'_i \cap S_j)$ could be recursively computed on table $S$. After that, we easily obtain the probability $\Pr\nolimits_{T \bowtie S}(Q_S \cap L_i)$.

In similar, for the probability $\Pr_{T \bowtie S}(Q_T)$, we could use the column $F_{T \to S}$ to correct the PDF from table $T$ to $T \bowtie S$. We assume that the  domain space $\Dom(T)$ are divided into regular sub-space $T_1, T_2, \dots T_m$ s.t. all attributes of $T$ are locally independent of $F_{T \to S}$ in each $T_k$. That is, for  for any space $Q' \in \Dom(T)$, we have 
\begin{equation*}
\Pr\nolimits_{T}((Q' \cap T_k) \wedge F_{T \to S} = f) = \Pr\nolimits_{T[T_k]}(Q' \cap T_k)  \cdot \Pr\nolimits_{T[T_k]}(F_{T \to S} = f).
\end{equation*}
We also have
\begin{equation}
\label{eq: Tsplit}
\begin{split}
& \Pr\nolimits_{T \bowtie S}(Q_T) \\
& = \frac{|T|}{|T \bowtie S|} \cdot 
\sum_{k} \sum_{f} \Pr\nolimits_{T}((Q_T \cap T_k) \wedge F_{T \to S} = f) \cdot f \\
& = \frac{|T|}{|T \bowtie S|} \cdot 
\sum_{k} \sum_{f} \Pr\nolimits_{T[T_k]}(Q_T \cap T_k)  \cdot \Pr\nolimits_{T[T_k]}(F_{T \to S} = f) \cdot f \\
& = \frac{|T|}{|T \bowtie S|} \cdot 
\sum_{k} \left( \Pr\nolimits_{T[T_k]}(Q_T \cap T_k) \cdot \sum_{f} \Pr\nolimits_{T[T_k]}(F_{T \to S} = f) \cdot f \right) \\
& = \frac{|T|}{|T \bowtie S|} \cdot 
\sum_{k} \Pr\nolimits_{T[T_k]}(Q_T \cap T_k) \cdot \mathbb{E}_{T[T_k]}[F_{T \to S}],
\end{split}
\end{equation}
where $\mathbb{E}_{T[T_k]}[F_{T \to S}]$ denotes the expected value of $F_{T \to S}$ restricted to all tuples of $T$ in the sub-domain $T_k$. All probabilities $\Pr\nolimits_{T}(Q_T \cap T_k)$ could be recursively obtained on table $T$. 

Using the join-key frequency local independence, we could obtain the probability $\Pr\nolimits_{T \bowtie S}(Q_S \cap L_i)$ over table $S$ for each $i$ and $\Pr\nolimits_{T \bowtie S}(Q_T)$ from table $T$ individually. Putting Eq.~\eqref{eq: Ssplit} and Eq.~\eqref{eq: Tsplit} into Eq.~\eqref{eq: TSsplit}, we easily obtain the probability of $Q$ on table $T \bowtie S$.

\smallskip
\noindent \underline{\textbf{Step~3: Recursive processing.}}
By Step~2, we need to further obtain each probability $\Pr\nolimits_{S}(Q'_i \cap S_j)$ from $S$ and $\Pr\nolimits_{T}(Q_T \cap T_k)$ from $T$. If $S$ (or $T$) is a single table, the probability could be directly computed using the underlying model of $\Pr_{S}(B_1, B_2, \dots, B_n)$. In \Glue, we have no limitation on how the model is built. It could either be a simple histogram or a complex SPN model. The probability could even be set to some magic number in some applications. This reserves great flexibility to the underlying DBMS to choose different \CE algorithms to meet different requirements, i.e., much faster for OLTP or more accurate for OLAP. Even more, \Glue also allows the QO to use different \CE methods on different tables in a database, so users could specify the most suitable \CE method according to the statistical characteristics of each table. 

If $S$ (or $T$) is not a single table, we could recursively decompose its joined tables for probability computation. Let $\Pr\nolimits_{S}(Q'_S) = \Pr\nolimits_{S}(Q_S \cap L_i \cap S_j)$ denote each query on a sub-range of $Q_S$ and $S = U \bowtie V$. We further divide $\Dom(S) = \Dom(U) \times \Dom(V)$ into regular sub-domains $L'_1, L'_2, \dots, L'_p$ in terms of $V$ s.t. attributes $U$ and $V$ are independent in $L'_i$, i.e., i.e., the cross-table local independence holds. 
Meanwhile, we divide the domain space $\Dom(V)$ to regular sub-space $V_1, V_2, \dots V_c$ s.t. all attributes of $V$ are locally independent of $F_{V \to U}$ in each $V_j$, i.e., the join-key frequency local independence holds. After that, we need to compute $\Pr\nolimits_{S}(Q'_S \cap L'_i \cap V_j)$ for all $1 \leq i \leq p$ and $1 \leq j \leq c$. The decomposition processing of table $T$ is similar. We iterate until both $U$ and $V$ are single tables. Notice that, we could organize all tables as a tree structure based on their decomposition manner, where each leaf node is a single table and each inner node is a join table. We defer the details on how to construct this decomposition tree in next section.

\subsection{\Glue for \CE}
\label{sec: method-3}

We  show how \Glue could be utilized for \CE. We first introduce the basic algorithm to process a single query, and then present how \Glue used in the plan search process of QO.

\smallskip
\noindent \underline{\textbf{Basic \CE algorithm in \Glue.}}
We present the complete algorithm \textsc{\Glue-CardEst} for \CE on a single query using \Glue in Figure~\ref{fig: algogluecardest}. We assume that the join tree has already been constructed offline with the root node $N$ and push the query $Q$ onto $N$. Basically, if $N$ is a leaf node on single table, we fetch and return the probability of $Q$ from the underlying model over $N$.
Otherwise, we fetch the cross-table local independence division conditions and split $Q$ to $Q_T$ and multiple $Q_S \cap L_j$ on the left and right child, respectively. If $Q_T$ (or $Q_S$) is empty, this indicates we have no predicates constraints over attributes in table $T$. We directly return $1$ to be the probability of $Q_T$ on node $N$. Else, for both $Q_T$ and $Q_S \cap L_j$, we divide it into many sub-ranges $Q'_T$ and $Q'_S$ by intersecting with each $T_k$ and $S_j$, respectively. The probability of $Q'_T$ and $Q'_S$ are obtained by recursively calling \textsc{\Glue-CardEst} on the sub-tree rooted at node $T$ and $S$, respectively. After collecting them, we compute the probability of $Q_T$ and $Q_S \cap L_j$ using Eq.~\eqref{eq: Tsplit} and Eq.~\eqref{eq: Ssplit}, respectively. They are merged together by Eq.~\eqref{eq: TSsplit} to obtain the cardinality of $Q$ on $T \bowtie S$. 

\smallskip
\noindent \underline{\textbf{Time complexity analysis.}}
Let $h$ be the height of the decomposition tree and $\ell$ be the maximum number of sub-space split in each node. Assume that the probability could be obtained in $O(t)$ time on each leaf node and $O(1)$ on each inner node. The time cost of \textsc{\Glue-CardEst} is $O(\ell^ht)$ as there exists at most $O(\ell^h)$ sub-space to be computed. This cost is affordable in nowdays DBMS due to the following reasons:
1) $O(t)$ is often very low on single tables using simple models such as histogram or ML models such as SPN~\cite{hilprecht2019deepdb} or FSPN~\cite{zhu2020flat};
2) $h$ is less than the number of joined tables in the DBMS, which is often a small constant number. We could also tune the tree construction method to decrease $h$;
and 3) the computation of different sub-space is easy to do in parallel.

\begin{figure}[t]
	\rule{\linewidth}{1pt}
	\leftline{~~~~\textbf{Algorithm} \textsc{\Glue-CardEst$(N, Q)$}}
	\vspace{-1em}
	\begin{algorithmic}[1]
		\IF{$N$ is a leaf node on single table $T$}
			\RETURN $\Pr_{T}(Q)$ from the underlying \CE model on $T$
		\ENDIF
		\STATE $Q_S, Q_T \gets Q$
		\IF{$Q_T$ is empty}
			\STATE $\Pr\nolimits_{T \bowtie S}(Q_T) \gets 1$
		\ENDIF
		\IF{$Q_S$ is empty}
		\STATE $\Pr\nolimits_{T \bowtie S}(Q_S) \gets 1$
		\ENDIF
		\FOR{each $T_k$ of $T$}
			\STATE $Q'_T \gets Q_T \cap T_k$
			\STATE $\Pr\nolimits_{T}(Q'_S) \gets$  \textsc{\Glue-CardEst}$(T, Q'_T)$
		\ENDFOR
		\STATE compute $\Pr\nolimits_{T \bowtie S}(Q_T)$ by Eq.~\eqref{eq: Tsplit} 
		\FOR{each $L_i$ of $S$}
			\FOR{each $S_j$ of $S$}
				\STATE $Q'_S \gets Q_S \cap L_i \cap S_j$
				\STATE $\Pr\nolimits_{S}(Q'_S) \gets$  \textsc{\Glue-CardEst}$(S, Q'_S)$
			\ENDFOR
			\STATE compute $\Pr\nolimits_{T \bowtie S}(Q_S \cap L_i)$ by Eq.~\eqref{eq: Ssplit} 
		\ENDFOR
		\STATE compute $\Pr\nolimits_{T \bowtie S}(Q)$ by Eq.~\eqref{eq: TSsplit} 
		\RETURN $\Pr_{T \bowtie S}(Q) \cdot |T \bowtie S|$
	\end{algorithmic}
	\rule{\linewidth}{1pt}
	\caption{\textsc{\Glue-CardEst} Algorithm.}
	\label{fig: algogluecardest}
\end{figure}

\smallskip
\noindent \underline{\textbf{\Glue in Plan Search Process.}}
Next, we show how \Glue could be seamlessly used in the plan search process. Unlike with existing \CE methods, such as~\cite{zhu2020flat, yang2019deep, yang2020neurocard, hilprecht2019deepdb}, which compute the cardinality for each sub-plan query from scratch, \Glue done in a more elegant manner. We utilize the overlaps among different sub-plan queries to avoid redundant computation. 

Given the join decomposition tree, for each node $N$, we associate it with a set $Cd(N)$. Each element in $Cd(N)$ is a pair $(D, p)$ where $D$ is a sub-domain space and $q$ is its probability. In the first traversal, we compute the probability of query $Q$ itself. The domain space is recursively divided into multiple sub-space in each node, we store each sub-space and its related probability on each node $N$ to obtain $Cd(N)$. Next time, for any sub-plan query $Q'$ of $Q$,
we do not need to decompose and compute the range of $Q'$ on the overlapped parts w.r.t.~$Q$. For non-overlapped sub-trees, which indicate $Q'$ does not touch the tables within this sub-tree, the sub-tree root node just returns $1$ as the probability on this branch.


\section{\Glue Construction}
\label{sec: struc}

In this section, we discuss how to construct the \Glue structure. 
We first introduce the details on space division to derive the local independence (in Section~\ref{sec: struc-1}) and then outline the method for building join decomposition tree (in Section~\ref{sec: struc-2}).

\subsection{Space Division Method}
\label{sec: struc-1}

We present the details on how to perform the space division in cross-table and join-key frequency local independence. We continue with the join table node $T \bowtie S$ with the left child $T$ and right child $S$. Assume that we have obtained the set of samples $\Smp(T), \Smp(S), \Smp(T \bowtie S)$ following the distribution of $T$, $S$ and $T \bowtie S$, respectively. Note that $\Smp(T \bowtie S)$ is not obtained by joining $\Smp(T)$ and $\Smp(S)$ directly, which does not obey the distribution of $T \bowtie S$. We could sample them individually using some methods such as~\cite{zhao2018random,li2016wander}. In these samples, we assume that the fanout columns $F_{S \to T}$ and $F_{T \to S}$ are given, which could be easily obtained by scanning all sampled tuples once. Then, we derive the space division conditions using these samples. 

According to the \Glue probability computation process in Section~\ref{sec: method}, the left child node $T$ only needs to divide the domain space $\Dom(T)$ to decouple $F_{T \to S}$with attributes $\{A_i\}$ of $T$. We could compute the pairwise correlation value $s(A_i, F_{T \to S})$ over $\Smp(T)$, such as RDC score~\cite{lopez2013randomized}, between each attribute $A_i$ of $T$ and $F_{T \to S}$. If some attributes $A_i$ are highly correlated with $F_{T \to S}$, i.e., $s(A_i, F_{T \to S})$ is larger than a threshold, we could randomly split the domain $\Dom(A_i)$ into several parts. This could be done for an evenly splitting of continuous attributes or randomly assignment of categorical attributes. Intuitively, after division, all tuples in the same part tends to have similar values in terms of $A_i$, so $A_i$ is not easily affected by $F_{T \to S}$ and more likely to be mutually independent. We iterate this process until all attributes $A_i$ are not highly correlated with $F_{T \to S}$. After that, for each divided domain $T_k$, we compute the estimated value $\widehat{\mathbb{E}}_{T[T_k]}[F_{T \to S}]$ of $F_{T \to S}$ over $\Smp(T)$ and store it for probability computation.

Similarly, for the right child $S$, we first compute the pairwise correlations between $s(A_i, B_j)$ over $\Smp(T \bowtie S)$, select an attribute $B_j$ maximizing the score $s(A_i, B_j)$ and perform the space division. After that, we once again divide $\Dom(S)$ over $\Smp(S)$ to decouple $\{B_j\}$ with $F_{S \to T}$. As these division conditions are all over $\Dom(S)$, we could merge them together afterwards. Sometimes, the two division steps over $S$ could also be do altogether, i.e., we choose to divide the attribute $B_j$ maximizes $\min (s(A_i, B_j), s(B_j, F_{S \to T}))$ each time. After computation, we also compute and store the estimated value $\widehat{\mathbb{E}}_{S[S_j]}[F_{S \to T}]$ of $F_{S \to T}$ over $\Smp(S)$ for each region $S_k$.

It is worth mentioning that our division strategy is only a heuristic rule, which has been shown to perform well in~\cite{zhu2020flat}. Essentially, \Glue is open for any space division method, such as grid-based clustering, as long as it could generate regular sub-space. Finding the suitable division method would be application-aware and an interesting future research work.

\subsection{Join Decomposition Tree}
\label{sec: struc-2}

On a higher perspective, \Glue is able to support any join schema, including but not limited to chain join, star join, cyclic join and self-join between tables and the complex mixture of them. 
We model the join schema graph among tables in a database as a graph $\mathcal{G} = (\mathcal{V}, \mathcal{E})$, where each node $T \in \mathcal{V}$ is a single relation table and each edge $(T, S) \in \mathcal{E}$ indicating $T$ could join with $S$. Each join relation could be: 
1) an inner join, as we explained the main idea of \Glue in Section~\ref{sec: method}, or outer, left or right join;
2) a one-to-many PK-FK join or a many-to-many FK-FK join;
and 3) an equal join or even unequal join.

A decomposition tree $\mathcal{T}$ is valid if and only if:
1) the root node is full join of all tables in $\mathcal{V}$;
2) each inner node $N$ splitting its join table $S \bowtie T$ to left child $S$ and right child $T$ s.t.~tables in $S$ and $T$ form connected components in $\mathcal{G}$ with internal join edges;
and 3) each leaf node corresponds to a single table in $\mathcal{V}$.
Obviously, each decomposition tree corresponds to a plan tree for a query touching all tables in $\mathcal{V}$. Therefore, it could be generated using a similar method for plan generation. 

In general, we could apply a dynamic programming method to construct the tree $\mathcal{T}$. Each time on a node $N$, we split its tables to $S$ joining with $T$ such that this decomposition minimizes $Cost(S) + Cost(T) + Cost(S, T)$. Here $Cost(S)$ (or $Cost(T)$) defines the recursively defined cost in terms of \CE task over underlying node $S$ (or $T$). $Cost(S, T)$ describes the cost for combining $S$ and $T$ together. Unlike with the plan generation with a specified cost model, the case for our \CE is a bit more complex. The cost model needs to consider the following aspects:

1) the sampling cost. As stated in Section~\ref{sec: method-1}, we need to apply samples on $T$, $S$ and $T \bowtie S$ to obtain the division conditions and expected value. Obviously, the sampling could be done easily when $T$ or $S$ contains less number of tables. For example, if we restrict $S$ to be a single table each time, in similar to the left-deep plan restriction, we could easily obtain the samples on $T \bowtie S$ by sampling tuples in $S$ w.r.t.~samples in $T$ using Olken's sampling algorithm. If $T$ or $S$ contains multiple tables, we may need to sample for $T \bowtie S$ individually and can not reuse the existing samples. We could measure this by $\min(||S||, ||T||)$, where $||T||$ denote the number of tables in $T$.

2) the error cost. The correlations between different pairs of $T$ and $S$ are different, so as the decomposition error. Intuitively, the smaller the correlation scores between  $T$ and $S$, the easier of the local independence exists. We could measure this by $s(T, S) = \max_{i, j}(s(A_i, B_j))$ where $s(A_i, B_j)$ is the correlation score between attributes $A_i$ of $T$ and $B_j$ of $S$.

3) the inference cost. By the time complexity analysis in Section~\ref{sec: method-2}, the inference cost is determined by the tree height $h$ and sub-space division number $l$. Obviously, $h$ is lower when $T$ and $S$ have balanced number of tables, and $l$ is lower when the correlations between $T$ and $S$ have higher correlations. Therefore, we could measure this by $s(T, S)^{\max(||S||, ||T||)}$. 

4) the modeling cost. This refers to the base cost of building \CE methods on each single table $T$. We denote it as a function $g(T)$, which is related to the structure learning time complexity of the underlying \CE algorithm. For histogram, it is linear w.r.t.~the number of attributes in $T$. For SPN/FSPN and BN, it is polynomial and exponential w.r.t.~the number of attributes in $T$, respectively.

Putting them together, we obtain the following cost model function
\begin{equation}
\label{eq: cost}
\begin{split}
Cost(S \bowtie T) & = \alpha \min(||S||, ||T||) + \beta s(T, S) \\
& + \gamma s(T, S)^{\max(||S||, ||T||)}  + Cost(S) + Cost(T), \\
Cost(T) & = g(T),
\end{split}
\end{equation}
where $\alpha$, $\beta$ and $\gamma$ are all hyper-parameters tuning the weights of each part. Some aspects, such as the sampling cost and inference cost, are conflict with each other. We could emphasize different parts in different scenarios, e.g., fast inference for OLTP or low error for OLAP. Note that, we measure each aspect using the most straightforward metric. \Glue is open for any complex cost model and it is also an interesting future research work.

Using Eq.~\eqref{eq: cost}, we could apply the dynamic programming method to construct the decomposition tree. The procedures are similar to plan generation. We omit it for simplicity. We could also apply some heuristic rules, such as greedy search, to find near-optimal result. Moreover, we do not restrict to build only one join decomposition tree for a database. We could build each for each frequently occurred join schema in the query workload.

\smallskip
\noindent \underline{\textbf{Model update.}}
Conceptually, the join decomposition tree is independent of the underlying models on single tables. Thus, they could be updated individually. When data changes on some tables, the corresponding \CE models are updated accordingly. For the join decomposition tree, it is more robust for data changes. We could periodically examine whether the local independence still holds in each sub-space. If not, we re-split the sub-space accordingly.


\section{\Glue for Distinct Count}
\label{sec: distinct}

In this section, we discuss how to adapt \Glue to count distinct number of values, which is frequently occurred in SQL queries with \textsf{distinct} predicate. We first show how to adapt existing \CE model to support distinct count on single table, and then how the framework in \Glue could support join queries. 

Following Section~\ref{sec: prelim}, from a statistical perspective, the distinct count of query $Q$ on table $T$ could de defined as $\Dis(T, Q) =|\{q \in Q| \Pr_{T}(q) > 0\}|$. That is, each point $q$ that can occur in the space of $Q$ is counted exactly once. Recall that, $T$ could be either single relational table or join table. In traditional methods, histogram and multiple sampling-based \CE methods could be used to count distinct number of values. For ML-based methods, we find that the SPN model~\cite{poon2011sum} and FSPN model~\cite{wu2020fspn} could also support distinct count with small adaptions. We elaborate the details as follows.

\smallskip
\noindent \underline{\textbf{Distinct count on single table.}}
For the SPN model, it models the joint PDF $Pr_{T}(A_1, A_2, \dots, A_k)$ using sum and product operations. Each sum node decomposes the joint PDF into weight sum of smaller models and each product node find local independence among different groups of attributes. Each leaf node in SPN maintains a histogram over a singleton attribute on some data.The distinct count could be done in similar to its probability inference as follows:

1) on leaf node modeling $Pr_{T'}(A_i)$, we easily obtain $\Dis(T', Q_i)$ by scanning the histogram of $A_i$ in $Q_i$'s range and send it to its father node.

2) on sum node modeling $Pr_{T}(A') = \sum_i w_i Pr_{T_i}(A')$, we restrict that each sum node divides the domain space into non-overlapping regular sub-space and each $T_i$ contains all tuples in a sub-space. As a results, we easily sum $\Dis(T, Q') = \sum_i \Dis(T_i, Q')$ from all children.

3) on product node modeling $Pr_{T'}(A') = \prod_j Pr_{T'}(A'_j)$, since $A'_j$ are all mutually independent, we know $Pr_{T'}(A') > 0$ if and only if $Pr_{T'}(A'_j) > 0$ for all $j$. That is, each distinct value of $A'$ must be counted in each $A_j$. Therefore, we easily have $\Dis(T', Q_{A'}) = \prod_j \Dis(T', Q_{A'_j})$.

The distinct count on SPN is the same as probability inference, which is linear w.r.t.~its node size. For FSPN, the method is similar as long as we count and add the distinct number in each sub-space specified by its factorize and split nodes.

\smallskip
\noindent \underline{\textbf{Distinct count in \Glue.}}
Following the local independence space decomposition in Section~\ref{sec: method-2}, the distinct counting in \Glue is also straightforward. First, Eq.~\eqref{eq: TSsplit} split the range of $Q$ to $Q_T$ and each $Q_S \cap L_j$. Due to the cross-table local independence, we obtain 
\begin{equation}
\label{eq: TSsplit-Dis}
\Dis(T \bowtie S, Q) = \sum_i \Dis(T \bowtie S, Q_T) \cdot \Dis(T \bowtie S, Q_S \cap L_i).
\end{equation}
For $\Dis(T \bowtie S, Q_T)$, we could obtain it from the joint PDF over $T$. By Eq.~\eqref{eq: Tsplit}, we have 
\begin{equation}
\label{eq: Tsplit-Dis}
\Dis(T \bowtie S, Q_T) = \sum_k 
\begin{cases}
\Dis(T, Q_T \cap T_k), & \text{if } \mathbb{E}_{T[T_k]}[F_{T \to S}] > 0; \\
0, & \text{if } \mathbb{E}_{T[T_k]}[F_{T \to S}]  = 0.
\end{cases}
\end{equation}
This is because in the sub-domain of each $T_k$, all attributes of $T$ are independent of $F_{T \to S}$ due to join-key frequency local independence. $\mathbb{E}_{T[T_k]}[F_{T \to S}]  = 0.$ indicates $\Pr_{T}(Q_T \wedge F_{T \to S} = f) = 0$ for all $f$, so no values in this sub-domain would be scattered from $T$ to $T \bowtie S$. Otherwise, $\mathbb{E}_{T[T_k]}[F_{T \to S}] > 0$ indicates $\Pr_{T}(Q_T \wedge F_{T \to S} = f) > 0$ for some $f$, so a value in $Q_T$ would occur in $T \bowtie S$ as long as it occurs in $T$. For each $\Dis(T, Q_T \cap T_k)$, it could be recursively obtained from table $T$. If $T$ is a single table, we return the distinct count value using the algorithm on single table. 

Similarly, for each $\Dis(T \bowtie S, Q_S \cap L_i)$, by Eq.~\eqref{eq: Ssplit}, we have 
\begin{equation}
\label{eq: Ssplit-Dis}
\begin{split}
& \Dis(T \bowtie S, Q_S \cap L_i) \\
& =
\sum_j 
\begin{cases}
\Dis(S, Q_S \cap L_i \cap S_j), & \text{if  } \mathbb{E}_{S[S_j]}[F_{S \to T}] > 0; \\
0, & \text{if  } \mathbb{E}_{S[S_j]}[F_{S \to T}] = 0,
\end{cases}
\end{split}
\end{equation}
where each $\Dis(S, Q_S \cap L_i \cap S_j)$ could be recursively obtained from table $S$. 

As a result, we could use the same framework of \textsc{\Glue-CardEst} for \CE to count distinct value with some small modifications:
1) replacing Eq.~\eqref{eq: TSsplit}, Eq.~\eqref{eq: Ssplit}  and Eq.~\eqref{eq: Tsplit} to Eq.~\eqref{eq: TSsplit-Dis}, Eq.~\eqref{eq: Ssplit-Dis}  and Eq.~\eqref{eq: Tsplit-Dis}, respectively;
and 2) in the base case, returning the distinct count value over a single table.

\section{Conclusions}

We propose \Glue, a general \CE framework that is able to merge single table \CE results to predict join query size. \Glue is flexible to support any underlying \CE method on single table and could steer to optimize different criteria. It is more flexible and adaptive to different datasets and query workloads, thus more suitable for deployment in real-world DBMS.


\bibliographystyle{ACM-Reference-Format}
\bibliography{main}

\end{document}